\documentstyle[12pt]{amsart}

\newcommand{\nc}{\newcommand}

\nc{\bSt}{\mbox{\bf{St}}}
\nc{\card}{\operatorname{card}}
\nc{\cd}{\operatorname{cd}}
\nc{\Ch}{\operatorname{Ch}}
\nc{\CHom}{\cal{H}om}
\nc{\codim}{\operatorname{codim}}
\nc{\Cone}{\operatorname{Cone}}
\nc{\dirlim}{\underset{\rightarrow}{\operatorname{lim}}}
\nc{\emp}{\emptyset}
\nc{\Fac}{\cal{F}ac}
\nc{\Hom}{\operatorname{Hom}}
\nc{\Id}{\operatorname{Id}}
\nc{\Ima}{\operatorname{Im}}
\nc{\Ind}{\operatorname{Ind}}
\nc{\invlim}{\underset{\leftarrow}{\operatorname{lim}}}
\nc{\Ker}{\operatorname{Ker}}
\nc{\Ob}{\operatorname{Ob}}
\nc{\one}{\mbox{\bf{1}}}
\nc{\Or}{\cal{O}r}
\nc{\Part}{\cal{P}art}
\nc{\sgn}{\operatorname{sgn}}
\nc{\Sh}{\cal{S}h}
\nc{\Tor}{\operatorname{Tor}}
\nc{\Vect}{\cal{V}ect}

\nc{\BA}{\Bbb A}
\nc{\ba}{\mbox{\bf{a}}}
\nc{\baJ}{\bar{J}}
\nc{\BAO}{\overset{\circ}{\Bbb A}}
\nc{\BB}{\Bbb B}
\nc{\BC}{\Bbb C}
\nc{\BE}{\overline{E}}
\nc{\BF}{\overline{F}}
\nc{\bF}{\mbox{\bf{F}}}
\nc{\bL}{\mbox{\bf{L}}}
\nc{\bM}{\mbox{\bf{M}}}
\nc{\bmu}{\vec{\mu}}
\nc{\BN}{\Bbb N}
\nc{\bnu}{\vec{\nu}}
\nc{\BP}{\Bbb P}
\nc{\bP}{\mbox{\bf{P}}}
\nc{\bq}{\mbox{\bf{q}}}
\nc{\BR}{\Bbb R}
\nc{\br}{\mbox{\bf{r}}}
\nc{\bs}{\mbox{\bf{s}}}
\nc{\bt}{\mbox{\bf{t}}}
\nc{\BUpsilon}{\bar{\Upsilon}}
\nc{\bw}{\mbox{\bf{w}}}
\nc{\bx}{\mbox{\bf{x}}}
\nc{\BZ}{\Bbb Z}
\nc{\bz}{\mbox{\bf{z}}}

\nc{\CA}{\cal A}
\nc{\CB}{\cal B}
\nc{\CC}{\cal C}
\nc{\CD}{\cal D}
\nc{\CE}{\cal E}
\nc{\CF}{\cal F}
\nc{\CH}{\cal H}
\nc{\CI}{\cal I}
\nc{\CJ}{\cal J}
\nc{\CK}{\cal K}
\nc{\CL}{\cal L}
\nc{\CM}{\cal M}
\nc{\CP}{\cal P}
\nc{\CS}{\cal S}
\nc{\CX}{\cal X}

\nc{\DO}{\overset{\circ}{D}}
\nc{\dpar}{\partial}

\nc{\fF}{\frak F}
\nc{\ff}{\frak f}
\nc{\fu}{\frak{u}}

\nc{\HO}{\overset{\circ}{H}}

\nc{\tc}{\tilde{c}}
\nc{\tE}{\tilde E}
\nc{\tF}{\tilde F}
\nc{\tfF}{\tilde{\frak{F}}}
\nc{\tJ}{\tilde{J}}
\nc{\tK}{\tilde K}
\nc{\tM}{\tilde{M}}
\nc{\tS}{\tilde S}
\nc{\ttheta}{\tilde{\theta}}
\nc{\tUpsilon}{\tilde{\Upsilon}}
\nc{\ty}{\tilde y}
\nc{\txi}{\tilde{\xi}}

\nc{\nen}{\newenvironment}
\nc{\ol}{\overline}
\nc{\ul}{\underline}
\nc{\ra}{\rightarrow}
\nc{\lra}{\longrightarrow}
\nc{\Lra}{\Longrightarrow}
\nc{\Lla}{\Longleftarrow}
\nc{\Llra}{\Longleftrightarrow}
\nc{\hra}{\hookrightarrow}
\nc{\iso}{\overset{\sim}{\lra}}

\nc{\Thm}[1]{Theorem~\ref{#1}}
\nc{\Prop}[1]{Proposition~\ref{#1}}
\nc{\Lem}[1]{Lemma~\ref{#1}}
\nc{\Cor}[1]{Corollary~\ref{#1}}
\nc{\Conj}[1]{Conjecture~\ref{#1}}
\nc{\Claim}[1]{Claim~\ref{#1}}
\nc{\Defn}[1]{Definition~\ref{#1}}
\nc{\Exa}[1]{Example~\ref{#1}}
\nc{\Rem}[1]{Remark~\ref{#1}}
\nc{\Note}[1]{Note~\ref{#1}}

\nen{thm}[1]{\label{#1}{\bf Theorem.\ } \em}{}
\nen{prop}[1]{\label{#1}{\bf Proposition.\ } \em}{}
\nen{lem}[1]{\label{#1}{\bf Lemma.\ } \em}{}
\nen{cor}[1]{\label{#1}{\bf Corollary.\ } \em}{}
\nen{conj}[1]{\label{#1}{\bf Conjecture.\ } \em}{}
\nen{claim}[1]{\label{#1}{\bf Claim.\ } \em}{}

\nen{defn}[1]{\label{#1}{\bf Definition.\ } }{}
\nen{exa}[1]{\label{#1}{\bf Example.\ } }{}

\nen{rem}[1]{\label{#1}{\em Remark.\ } }{}
\nen{note}[1]{\label{#1}{\em Note.\ } }{}
\nen{exer}[1]{\label{#1}{\em Exercise.\ } }{}

\begin{document}

%  top matter
\title[]{Localization of $\fu$-modules. I.\\ Intersection cohomology of real
arrangements}
\author{Michael Finkelberg}
\address{Independent Moscow University, 65-3 Mikloukho-Maklai St.,
apt. 86, Moscow 117342 Russia}
\author{Vadim Schechtman}
\address{Dept. of Mathematics, SUNY at Stony Brook, Stony Brook,
NY 11794-3651 USA}

\date{December 1994\\
hep-th/9411050}
\thanks{The second author was supported in part by NSF grant DMS-9202280}

\maketitle

\section{Introduction}

This paper is the first in a series. The main goal of the series
is to present a geometric construction of certain remarkable
tensor categories arising from quantum groups corresponding
to the value of deformation parameter $q$ equal to a root of unity,
cf. ~\cite{ajs}, these categories playing an important role in Lusztig program.

\subsection{} Let $\BA$ be a complex affine space, $\CH$ a finite collection
of hyperplanes in $\BA$. We suppose that all $H\in\CH$ are given by real
equations, and denote by $H_{\BR}\subset H$ the subspace of real points.
An arrangement $\CH$ induces naturally a stratification of
$\BA$ denoted by $\CS_{\CH}$ (for precise definitions see section 2 below).
The main goal of this paper is the study
of the category $\CM(\BA;\CS_{\CH})$ of perverse sheaves (of vector spaces
over a fixed ground field) over $\BA$ which are smooth along $\CS_{\CH}$.

In {\em section 1} we collect topological notations and known facts we will
need
in the sequel.

{\em Section 2.}
To each object $\CM\in\CM(\BA;\CS_{\CH})$ we assign "a linear algebra data".
Namely, for each {\em facet} of $\CH$ --- i.e. a connected component
of an intersection of some $H_{\BR}$'s --- we define a vector space
$\Phi_F(\CM)$; if $F$ is contained and has codimension $1$ in the closure of
another facet $E$, we define two linear operators between $\Phi_F(\CM)$ and
$\Phi_E(\CM)$ acting in opposite directions. These data contain all
information about $\CM$ we need.

In fact, it is natural to expect that the sheaf $\CM$ may be {\em uniquely
reconstructed} from the linear algebra data above (one can check this in
the simplest cases).

The spaces $\Phi_F(\CM)$ are defined using certain relative cohomology.
They are similar to a construction by R. MacPherson (see ~\cite{b2}).
The spaces $\Phi_F(\CM)$ are analogous to functors of vanishing cycles.

The main technical properties of functors $\CM\mapsto\Phi_F(\CM)$ are
(a) {\em commutation with Verdier duality}, and (b) {\em exactness}.
In fact, (b) follows from (a) without difficulty; the proof of (a)
is the principal geometrical result of this paper (see Theorem ~\ref{dual}).
Actually, we prove a more general statement concerning all complexes
smooth along $\CS_{\CH}$. This is a result of primary importance for us.

Using these linear algebra data, we define an exact functor from
$\CM(\BA;\CS_{\CH})$ to complexes of finite dimensional vector spaces
computing cohomology $R\Gamma(\BA;\bullet)$, see Theorem ~\ref{rgamma}.
This is the main result we will need below. The idea of constructing such
functorial complexes on categories of perverse sheaves
(quite analogous to the usual computation of singular cohomology using
cell decompositions) is due to A. Beilinson,
{}~\cite{b1}. It was an important source of inspiration for us.

A similar problem was considered in ~\cite{c}.

In {\em section 3} we present explicit computations for standard extensions
of an arbitrary one-dimensional local system over an open stratum.
The main result is the computation of {\em intersection cohomology}, see
Theorem ~\ref{inters}. In our computations we use some simple geometric
ideas from M. Salvetti's work, ~\cite{sa}. However, we do not use a more
difficult  main theorem of this paper. On the contrary, maybe our
considerations shed some new light on it.

\subsection{} The idea of this work appeared several years ago,
in some discussions with H\'{e}l\`{e}ne Esnault and Eckart Viehweg.
We use this occasion to express to them our sincere gratitude. We are also
very grateful to Paul Bressler for a stimulating discussion.
We thank Kari Vilonen for pointing out an error in an earlier version of the
manuscript.
We are greately
indebted to D. Kazhdan for his constant encouragement and attention to the
work.

\section{Topological preliminaries}

In this section we introduce our notations and recall some basic
facts from topology. Main references are ~\cite{ks}, \cite{bbd}.

\subsection{}
\label{derived} Throughout the paper, all our topological
spaces will be {\em locally compact}, in particular {\em Hausdorff}.
$\{ pt\}$ will denote a one-point space. For a topological space $X$,
$a_X:X\lra \{pt\}$ will be the unique map. If $Y\subset X$ is a subspace,
$\bar{Y}$ will denote the closure of $Y$.

Throughout this chapter we fix an arbitrary ground field $B$. {\em A vector
space} will mean a vector space over $B$. For a finite dimensional
vector space $V$, $V^*$ will denote a dual space.
$\Vect$ will denote a category
of vector spaces.

{\em A sheaf} will mean a sheaf of vector spaces. For a topological
space $X$, $\Sh(X)$ will denote a category of sheaves over $X$,
$\CD^*(X)$ will denote the derived category of $\Sh(X)$,
$*=+,-,b$ or $\emp$ will have the usual meaning.

For $p\in\Bbb Z$, $\CD^{\leq p}(X)\subset \CD(X)$ will denote
a full subcategory of complexes $\CK$ with $H^i(\CK)=0$ for all
$i>p$.

If $\CA,\CB$ are abelian categories, we will say that a left exact functor
$F:\CD^*(\CA)\lra\CD(\CB)$ ($*=+,-,b$ or $\emp$)
has {\em cohomological dimension $\leq r$}, and write
$\cd (F)\leq r$, if $H^i(F(A))=0$ for any $A\in\CA$ and $i>r$. (Left exactness
here means that $H^i(F(A))=0$ for $i<0$).

\subsection{} Let $f:X\lra Y$ be a continuous map. We will use the following
notations for standard functors.

$f^*:\CD(Y)\lra\CD(X)$ --- the inverse image;
$f_*:\CD^+(X)\lra\CD^+(Y)$ --- (the right derived of) the direct image;
$f_!:\CD^+(X)\lra\CD^+(Y)$ --- (the right derived of) the direct image
with proper supports;
$f^!:\CD^+(Y)\lra\CD^+(X)$ --- the right adjoint to $f_!$ (defined
when $f_!$ has finite cohomological dimension), see ~\cite{ks}, 3.1.

We will denote the corresponding functors on sheaves as follows:
$f^*:\Sh(Y)\lra\Sh(X);\ R^0f_*:\Sh(X)\lra\Sh(Y);\
R^0f_!:\Sh(X)\lra\Sh(Y)$.

We will denote $R\Gamma(X;\cdot):=p_{X*};\  R\Gamma_c(X;\cdot):=p_{X!}$.
For $\CK\in\CD^+(X),\ i\in\BZ$, we set
$H^i(X;\CK):=H^i(R\Gamma(X;\CK));\  H^i_c(X;\CK):=H^i(R\Gamma_c(X;\CK))$.

For $V\in\Vect=\Sh(\{ pt\})$ we denote by $V_X$ the constant sheaf $p^*_XV$.

If $Y$ is a subspace of $X$, we will use a notation $i_{Y,X}$,
or simply $i_Y$ if $X$ is understood, for the
embedding $Y\hra X$. If $Y$ is open in $X$, we will also write $j_Y$
instead of $i_Y$.

For $\CK\in\CD(X)$, we will use notations $\CK|_Y:=i^*_{Y,X}\CK$;
$R\Gamma(Y;\CK):=R\Gamma(Y;\CK|_Y),\ H^i(Y;\CK):=H^i(Y;\CK_Y)$,
etc.

\subsection{} We have functors
\begin{equation}
R\CHom :\CD^-(X)^{opp}\times\CD^+(X)\lra\CD^+(X),
\end{equation}
\begin{equation}
\otimes:\CD^-(X)\times\CD^*(X)\lra\CD^*(X)
\end{equation}
where $*=,-,b$ or $\emp$;
\begin{equation}
R\Hom=R\Gamma\circ R\CHom:\CD^-(X)^{opp}\times\CD^+(X)\lra\CD^+(\Vect)
\end{equation}
For $\CK,\CL\in\CD^b(X)$ we have
\begin{equation}
\Hom_{\CD(X)}(\CK,\CL)=H^0(R\Hom(\CK,\CL))
\end{equation}

We denote $\CK^*:=R\CHom(\CK,B_X)$.

\subsection{} Let $X$ be a topological space, $j:U\lra X$ an embedding
of an open subspace, $i: Y\lra X$ an embedding of the complement.
In this case $i_*=i_!$ and $j^!=j^*$. $R^0i_*$ and $R^0j_!$ are
exact, so we omit $R^0$ from their notation.
$j_!$ is the functor of extension by zero.
$i^!$ is the right derived of the functor $R^0i^!:\Sh(X)\lra \Sh(Y)$
of the subsheaf of sections with supports in $Y$ (in notations of ~\cite{ks},
$R^0i^!(F)=\Gamma_Y(F)$).

Let $\CK\in\CD^+(X)$. We will use the following notations for relative
cohomology: $R\Gamma(X,Y;\CK):=R\Gamma(X;j_!\CK|_U);\
H^i(X,Y;\CK):=H^i(R\Gamma(X,Y;\CK))$.

If $Z\hra Y$ is a closed subspace,
%$j_{Y-Z,Y}:Y-Z\hra Y$, $j_{X-Z,X}:X-Z\hra X$,
we have a canonical exact triangle
\begin{equation}
i_{X-Y!}\CK|_{X-Y}\lra i_{X-Z!}\CK|_{X-Z}
\lra i_{Y-Z!}\CK|_{Y-Z} \lra i_{X-Y!}\CK|_{X-Y}[1]
\end{equation}
(of course, $i_{X-Y}=j$)
inducing a long cohomology sequence
\begin{equation}
\label{cohom}
\ldots\lra H^i(X,Y;\CK)\lra H^i(X,Z;\CK)\lra H^i(Y,Z;\CK)
\overset{\dpar}{\lra} H^{i+1}(X,Y;\CK)\lra\ldots
\end{equation}

\subsection{} Let $X$ be a topological space such that the functor
$R\Gamma_c(X;\cdot)$ has a finite cohomological dimension.
We define {\em a dualizing complex}:
\begin{equation}
\omega_X:=a_X^!B\in\CD^b(X)
\end{equation}
For $\CK\in\CD^b(X)$ we set $D_X\CK=R\CHom(\CK,\omega_X)\in\CD^b(X)$.
If there is no risk of confusion we denote $D\CK$ instead of $D_X\CK$.
We get a functor
\begin{equation}
D:\CD^b(X)^{opp}\lra\CD^b(X)
\end{equation}
which comes together with a natural transformation
\begin{equation}
\Id_{\CD^b(X)}\lra D\circ D
\end{equation}

\subsection{Orientations} (Cf. ~\cite{ks}, 3.3.) Let $X$ be an $n$-dimensional
$C^0$-manifold.
We define an {\em orientation sheaf} $\Or_X$ as the sheaf associated to
a presheaf $U\mapsto \Hom_{\BZ}(H^n_c(U;\BZ),\BZ)$. It is a locally constant
sheaf of abelian groups of rank $1$. Its tensor square is constant.
By definition, {\em orientation of $X$} is an isomorphism
$\Or_X\iso\BZ_X$.

We have a canonical isomorphism
\begin{equation}
\omega_X\iso\Or_X\otimes_{\BZ}B[n]
\end{equation}

\subsection{Homology} Sometimes it is quite convenient to use
homological notations. Let $Y\subset X$ be a closed subspace of
a topological space $X$ ($Y$ may be empty), $\CK\in\CD^b(X)$.
We define homology groups as
$$
H_i(X,Y;\CF):=H^i(X,Y;\CF^*)^*.
$$
These groups behave covariantly with respect to continuous mappings.

Let $\sigma$ be {\em a relative singular $n$-cell}, i.e. a continuous
mapping
$$
\sigma: (D^n,\dpar D^n)\lra (X,Y)
$$
where, $D^n$ denotes a standard closed unit ball in $\BR^n$.
We supply $D^n$ with the standard orientation. Let $\DO^n$ denote the interior
of $D^n$.

Suppose that $\sigma^*\CK|_{\DO^n}$ is constant. Then by Poincar\'{e} duality
we have isomorphisms
$$
H_n(D^n,\dpar D^n;\sigma^*\CK)=H^n_c(\DO^n;\sigma^*\CK^*)^*\iso
H^0(\DO^n;\sigma^*\CK).
$$
(recall that $\DO^n$ is oriented).
Thus, given an element $s\in H^0(\DO^n;\CK)$, we can define a homology class
$$
cl(\sigma;s):=\sigma_*(s)\in H_n(X,Y;\CK).
$$
We will call the couple $(\sigma;s)$
{\em a relative singular $n$-cell for $\CK$}.

These classes enjoy the following properties.

\subsubsection{Homotopy} Let us call to cells $(\sigma_0;s_0)$
and $(\sigma_1,s_1)$ {\em homotopic} if there exists a map
$$
\sigma: (D^n\times I,\dpar D^n\times I)\lra (X,Y)
$$
(where $I$ denotes a unit interval) such that $\sigma^*\CK|_{\DO^n\times I}$
is constant,  and a section
$$
s\in H^0(\DO^n\times I;\sigma^*\CK)
$$
such that $(\sigma,s)$ restricted to $D^n\times\{ i\}$
is equal to $(\sigma_i;s_i),\ i=0,1$.

We have

(H) {\em if $(\sigma_0;s_0)$ is homotopic to $(\sigma_1;s_1)$ then
$cl(\sigma_0;s_0)=cl(\sigma_1;s_1)$.}

\subsubsection{Additivity} Suppose $D^n$ is represnted as a union of its
upper and lower half-balls $D^n=D^n_1\cup D^n_2$ where
$D^n_1=\{ (x_1,\ldots, x_n)\in D^n|\ x_n\geq 0\}$ and
$D^n_2=\{ (x_1,\ldots, x_n)\in D^n|\ x_n\leq 0\}$. Let us supply $D^n_i$
with the induced orientation.

Suppose we are given a relative $n$-cell $(\sigma;s)$ such that
$\sigma(D^n_1\cap D^n_2)\subset Y$.
Then its restriction to $D^n_i$ gives us two relative $n$-cells
$(\sigma_i;s_i)$,
$i=1,2$. We have

(A) $cl(\sigma;s)=cl(\sigma_1;s_1)+cl(\sigma_2;s_2)$.

\subsection{} We will call {\em a local system} a locally constant sheaf
of finite rank.

Let $X$ be a topological space, $i:Y\hra X$ a subspace. We will say that
$\CK\in\CD(X)$ is {\em smooth} along $Y$ if all cohomology sheaves
$H^p(i^*\CK),\ p\in\BZ,$ are local systems.

We will call a {\em stratification} of $X$ a partition $X=\bigcup_{S\in\CS}S$
into a finite disjoint of locally closed subspaces ---
{\em strata} --- such that the closure of each stratum is the union
of strata.

We say that $\CK\in\CD(X)$ is {\em smooth along $\CS$} if it is smooth
along each stratum. We will denote by $\CD^*(X;\CS)$ the full subcategory
of $\CD^*(X)$ ($*=+,-,b$ or $\emp$) consisting of complexes smooth along $\CS$.

\subsection{} Let us call a stratification $\CS$ of a topological space
$X$ {\em good} if the following conditions from ~\cite{bbd}, 2.1.13
hold.

(b) All strata are equidimensional topological varieties. If a stratum
$S$ lies in the closure of a stratum $T$, $\dim S<\dim T$.

(c) If $i:S\hra X$ is a stratum, the functor $i_*:\CD^+(S)\lra \CD^+(X)$
has a finite cohomological dimension. If $\CF\in\Sh(S)$ is a local system,
$i_*(\CF)$ is smooth along $\CS$.

Let $\CS$ be a good stratification such that all strata have even dimension.
We will denote by $\CM(X;\CS)\subset\CD^b(X;\CS)$ the category of
smooth along $\CS$
perverse sheaves corresponding to the middle perversity $p(S)=-\dim\ S/2$, cf.
{\em loc. cit.}

\subsection{} Let $X$ be a complex algebraic variety. Let us call
its stratification $\CS$ {\em algebraic} if all strata are algebraic
subvarieties. Following ~\cite{bbd}, 2.2.1, let us call a sheaf
$\CF\in\Sh(X)$ {\em constructible} if it is smooth along some algebraic
stratification. (According to Verdier, every algebraic stratification
admits a good refinement.) We denote by $\CD^b_c(X)$ a full subcategory
of $\CD^b(X)$ consisting of complexes with constructible cohomology.

We denote by $\CM(X)\subset\CD^b_c(X)$ the category of perverse sheaves
corresponding to the middle perversity. It is a filtered union of categories
$\CM(X;\CS)$ over all good algebraic stratifications $\CS$, cf. {\em loc.cit.}

%If $X$ and $Y$ are complex algebraic varieties, we will call a functor
%$F:\CD^b(X)\lra\CD^b(Y)$ {\em exact} if it carries $\CM(X)$ to
%$\CM(Y)$.

\subsection{} Let $X$ be a complex manifold, $f:X\lra \BC$ a holomorphic map.
Set $Y=f^{-1}(0),\ U=X-Y$, let $i:Y\hra X,\ j:U\hra X$ denote the
embeddings.

We define a functor of {\em nearby cycles}
\begin{equation}
\label{near}
\Psi_f:\CD^b(U)\lra\CD^b(Y)
\end{equation}
as $\Psi_f(\CK)=\psi_f j_*(\CK)[-1]$ where $\psi_f$ is defined in ~\cite{ks},
8.6.1.

We define a functor of {\em vanishing cycles}
\begin{equation}
\label{vanish}
\Phi_f:\CD^b(X)\lra\CD^b(Y)
\end{equation}
as $\phi_f$ from {\em loc. cit.}

\subsection{}
\label{limit}
\begin{lem}{} Let $X$ be topological space, $Y\hra X$ a closed subspace,
$\CF\in\Sh(X)$. Then natural maps
$$
\dirlim\ H^i(U;\CF)\lra H^i(Y;\CF),\ i\in\BZ,
$$
where $U$ ranges through all the open neighbourhoods of $U$, are
isomorphisms.
\end{lem}

{\bf Proof.} See ~\cite{ks}, 2.5.1, 2.6.9. $\Box$

\subsection{Conic sheaves}
\label{conic} (Cf. ~\cite{ks}, 3.7.) Let $\BR^{*+}$ denote
the multiplicative group of positive real numbers. Let $X$ be a topological
space endowed with an $\BR^{*+}$-action.

Following {\em loc.cit.}, we will call a sheaf $\CF$ over $X$ {\em conic}
(with respect to the given $\BR^{*+}$-action)
if its restriction to every $\BR^{*+}$-orbit is constant.
We will call a complex $\CK\in\CD(X)$ {\em conic} if all its cohomology
sheaves are conic.

We will denote by $\Sh_{\BR^{*+}}(X)\subset \Sh(X)$,
$\CD^*_{\BR^{*+}}(X)\subset\CD^*(X),\ *=b,+,-$ or $\emp$,
the full subcategories of conic objects.

\subsubsection{}
\label{conicl}
\begin{lem}{} Let $U\hra X$ be an open subset. Suppose that for every
$\BR^{*+}$-orbit $O\subset X$, $O\cap U$ is contractible (hence, non-empty).
Then for every conic $\CK\in\CD^+(X)$ the restriction morphism
$$
R\Gamma(X;\CK)\lra R\Gamma(U;\CK)
$$
is an isomorphism.
\end{lem}

{\bf Proof.} See {\em loc. cit.}, 3.7.3. $\Box$

\section{Vanishing cycles functors}
\label{vanishcycl}

\subsection{Arrangements} Below we use some terminology from ~\cite{sa}.
Let $\BA_{\BR}$ be a
real affine space $\BA_{\BR}$ of dimension $N$, and $\CH_{\BR}=\{ H_{\BR}\}$
a finite set of distinct real affine hyperplanes in $\BA_{\BR}$. Such a set
is called a {\em real arrangement}. We pick once and for all a square root
$i=\sqrt{-1}\in\BC$.

Let $\BA=\BA_{\BR}\otimes_{\BR}\BC$ denote the complexification of
$\BA_{\BR}$, and $\CH=\{ H\}$ where $H:=H_{\BR}\otimes_{\BR}\BC$.
(A finite set of complex hyperplanes in a complex affine space
will be called a {\em complex arrangement}).

We will say that $\CH$ is {\em central} if $\bigcap_{H\in\CH}H$ consists
of one point.

For a subset $\CK\subset \CH$ denote
$$
H_{\CK}=\bigcap_{H\in \CK}H;\ _{\CK}H=\bigcup _{H\in \CK}H,
$$
and
$$
H_{\BR,\CK}=\bigcap_{H\in \CK}H_{\BR},
\ _{\CK}H_{\BR}=\bigcup_{H\in \CK}H_{\BR}.
$$
The nonempty subspaces $H_{\CK}$ and $H_{\CK,\BR}$ are called complex
and real {\em edges} respectively. Set
$$
\HO_{\CK}=H_{\CK}-\cup L,
$$
the union over all the complex edges $L\subset H_{\CK},\ L\neq H_{\CK}$. We set
$\HO_{\BR,\CK}=H_{\BR,\CK}\cap \BA_{\BR}$.
Connected components of $\HO_{\BR,\CK}$ are called {\em facets} of $\CH_{\BR}$.
Facets of codimension $0$ (resp., $1$) are called {\em chambers}
(resp., {\em faces}).

Let us denote by $\CS_{\CH}$ a stratification of $\BA$ whose strata are
all non-empty $\HO_{\CK}$. We will denote by $\BAO_{\CH}$ a unique open
stratum
$$
\BAO_{\CH}=\HO_{\emp}=\BA- _{\CH}H.
$$

In this Section we will study categories of sheaves $\CD(\BA;\CS_{\CH})$
and $\CM(\BA;\CS_{\CH})$.

We will denote by $\CS_{\CH_{\BR}}$ a stratification of $\BA_{\BR}$ whose
strata are all facets. We set
$$
\BAO_{\CH,\BR}=\BA_{\BR}- _{\CH}H_{\BR}.
$$
It is a union of all chambers.

\subsection{Dual cells}
\label{salv}(cf. ~\cite{sa}).
Let us fix a point $^Fw$ on each facet $F$. We will call this set of points
$\bw =\{^Fw\}$ {\em marking} of our arrangement.

For two facets $F,\ E$ let us write $F<E$ if $F\subset\BE$ and
$\dim\ F<\dim\ E$.
We will say that $E$ is {\em adjacent} to $F$. We will denote by $\Ch (F)$
the set of all chambers adjacent to $F$.

Let us call a {\em flag} a sequence
of $q-p+1$ facets $\bF=(F_p<F_{p+1}<\ldots <F_q)$ with $\dim\ F_i=i$.
We say that $F_p$ is {\em the beginning} and $F_q$ {\em the end} of $\bF$.

Let us denote by $^{\bF}\Delta$ a closed $(q-p)$-symplex with vertices
$^{F_p}w,\ldots,^{F_q}w$. Evidently, $^{\bF}\Delta\subset\BF_q$.

Suppose we are given two facets $F_p<F_q$, $\dim F_i=i$. We will denote
$$
D_{F_p<F_q}=\bigcup\ ^{\bF}\Delta,
$$
the sum over all flags beginning at $F_p$ and ending at $F_q$. This is a
$(q-p)$-dimensional cell contained in $\BF_q$.

For a facet $F$ let us denote
$$
D_F=\bigcup_{C\in \Ch (F)}D_{F<C}
$$
We set
$$
S_F:=\bigcup_{E,C: F<E<C;}D_{E<C},
$$
the union over all facets $E$ and chambers $C$. The space $S_F$ is contained
in $D_F$ (in fact, in the defintion of $S_F$ it is enough to take the union
over all facets $E$ such that
$\dim\ E=\dim\ F+1$). If $q=\codim\ F$, then
$D_F$ is homeomorphic to a $q$-dimensional disc and $S_F$ --- to a $(q-1)$-
dimensional sphere, cf. ~\cite{sa}, Lemma 6. We denote $\DO_F:=D_F-S_F$.
We will call $D_F$ {\em a dual cell corresponding to $F$}.

We set $\DO_{F<C}:=D_{F<C}\cap \DO_F$.

\subsection{Generalized vanishing cycles} Let $\CK\in\CD^b(\BA;\CS_{\CH})$,
$F$ a $p$-dimensional facet. Let us introduce a complex
\begin{equation}
\label{vc}
\Phi_F(\CK):=R\Gamma(D_F,S_F;\CK)[-p]\in\CD^b(pt)
\end{equation}
This complex will be called {\em a complex of generalized vanishing
cycles of $\CK$ at a facet $F$}.

Formally, the definition of functor $\Phi_F$ depends
upon the choice of a marking $\bw$. However, functors defined using
two different markings are canonically isomorphic. This is evident.
Because of this, we omit markings from the notations.

\subsection{Transversal slices}
\label{slice} Let $F$ be a facet of dimension $p$ which
is a connected component of a real edge $M_{\BR}$ with the complexification
$M$.
Let us choose a real affine subspace $L_{\BR}$ of codimension $p$ transversal
to $F$ and passing through $^Fw$. Let $L$ be its complexification.

Let us consider a small disk $L_{\epsilon}\subset L$ with the centrum
at $^Fw=L\cap M$. We identify $L_{\epsilon}$ with an affine space
by dilatation. Our arrangement induces a central arrangement $\CH_L$
in $L_{\epsilon}$. Given  $\CK\in\CD^b(\BA;\CS_{\CH})$,
define $\CK_L:=i_{L_{\epsilon}}^*\CK[-p]$.

\subsubsection{} {\bf Lemma.} {\em We have a natural isomorphism
\begin{equation}
\label{slicedu}
D(\CK_L)\iso (D\CK)_L
\end{equation}}

{\bf Proof.} Consider an embedding of smooth complex manifolds
$i_L: L\lra \BA$. Let us consider the following complexes:
$\omega_{L/\BA}:=i^!_L B_{\BA}$ (cf. ~\cite{ks}, 3.1.16 (i)) and
$\Or_{L/\BA}:=\Or_{L}\otimes_{\BZ}\Or_{\BA}$.
We have canonical isomorphism
$$
\omega_{L/\BA}\iso\Or_{L/\BA}[-2p]\otimes B.
$$
The chosen orientation of $\BC$ enables us to identify
$\Or_{\BA}$ and $\Or_L$, and hence $\Or_{L/\BA}$ with constant sheaves;
consequently, we get an isomorphism
$$
\omega_{L/\BA}\iso B[-2p].
$$
The canonical map
\begin{equation}
\label{closed}
i^!_L\CK\lra i^*\CK\otimes\omega_{L/\BA}
\end{equation}
(cf. ~\cite{ks}, (3.1.6)) is an isomorphism since singularities of $\CK$
are transversal to $L$ (at least in the neighbourhood of $^Fw$).
Consequently we get an isomorphism $i^!_L\CK\iso i^*\CK_L[-2p]$.

Now we can compute:
$$
D(\CK_L)=D(i_L^*\CK[-p])\iso D(i^*_L\CK)[2p]\iso i^!_LD\CK[p]
\iso i^*_LD\CK[-p]=(D\CK)_L,
$$
QED. $\Box$

\subsubsection{} {\bf Lemma.} {\em We have a natural
isomorphism
\begin{equation}
\label{sliceiso}
\Phi_{F}(\CK)\cong\Phi_{\{^Fw\} }(\CK_L)
\end{equation}}

This follows directly from the definition of functors $\Phi$.

This remark is often useful for reducing the study of functors $\Phi_F$ to
the case of a central arrangement.

\subsubsection{} {\bf Lemma.} {If $\CM\in\CM(\BA;\CS_{\CH})$
then $\CM_L\in\CM(L_{\epsilon};\CS_{\CH_{L_{\epsilon}}})$.}

{\bf Proof} follows from transversality of $L_{\epsilon}$ to singularities
of $\CM$. $\Box$

\subsection{Duality}
\label{dual}
\begin{thm}{} Functor $\Phi_F$ commutes with Verdier duality.
More precisely, for every $\CK\in\CD^b(\BA,\CS_{\CH})$ there exists a natural
isomorphism
\begin{equation}
\label{duala}
\Phi_F(D\CK)\iso D\Phi_F(\CK)
\end{equation}
\end{thm}

This is the basic property of our functors.

\subsection{$1$-dimensional case}
\label{one-dim} To start with the proof, let us treat first the simplest case.
Consider an arrangement consisting of one point --- origin --- in
a one-dimensional
space $\BA$. It has one $0$-face $F$ and two $1$-faces $E_{\pm}$ --- real rays
$\BR_{>0}$ and $\BR_{<0}$. A marking consists of two points
$w_{\pm}\in E_{\pm}$ and $F$. We set $w_{\pm}=\pm 1$ (we pick a coordinate on
$\BA$).

For a positive
$r$ denote $\BA_{\leq r}:=\{ z\in\BA|\ |z|\leq r\},\ S_r:=\dpar\BA_{\leq r}$;
$\BA_{< r},\ \BA_{\geq r}$, etc. have an evident meaning. We also set
$\BA_{(r',r'')}:=\BA_{>r'}\cap \BA_{<r''}.$ A subscript $\BR$
will denote an intersection of these subsets with $\BA_{\BR}$.
Evidently, $D_F=\BA_{\leq 1,\BR},\ S_F=S_{1,\BR}$. Define $D_F^{opp}:=
\BA_{\geq 1,\BR},\ Y:=i\cdot D_F^{opp}$.

One sees easily (cf. {\em infra}, Lemma ~\ref{isom}) that one has isomorphisms
\begin{equation}
\label{isom-one}
\Phi_F(\CK)\cong R\Gamma(\BA,S_F;\CK)\iso R\Gamma(\BA,D_F^{opp};\CK).
\end{equation}

Let us choose real numbers $\epsilon, r', r''$ such that $0<\epsilon<r'<1<r''$.
Set $Y:=\epsilon i\cdot D_F^{opp}$. Denote $j:=j_{\BA-S_F}$.
We have natural isomorphisms
$$
D\Phi_{F}(\CK)\cong DR\Gamma(\BA,S_{F};\CK)
\cong R\Gamma_c(\BA; j_{*}j^*D\CK)
$$
(Poincar\'{e} duality)
$$
\cong R\Gamma(\BA,\BA_{\geq r''};j_{*}j^*D\CK)\cong
R\Gamma(\BA,Y\cup \BA_{\geq r''};j_{*}j^*D\CK)
$$
(homotopy).
Consider the restriction map
\begin{equation}
res: R\Gamma(\BA,Y\cup \BA_{\geq r''};j_*j^*D\CK)\lra
R\Gamma(\BA_{\leq r'},Y\cap \BA_{\leq r'};D\CK)
\end{equation}
We claim that $res$ is an isomorphism.
In fact, $\Cone (res)$ is isomorphic to
\begin{eqnarray}
R\Gamma(\BA,\BA_{\leq r'}\cup\BA_{\geq r''}\cup Y;j_*j^*D\CK)=
R\Gamma_c(\BA_{< r''},\BA_{\leq r'}\cup Y;j_*j^*D\CK)\cong\nonumber\\
\cong DR\Gamma(\BA_{< r''}-(\BA_{\leq r'}\cup Y);j_!j^*\CK)\nonumber
\end{eqnarray}
We have by definition
$$
R\Gamma(\BA_{< r''}-(\BA_{\leq r'}\cup Y);j_!j^*\CK)=
R\Gamma(\BA_{(r',r'')}-Y,S_F;\CK)
$$
On the other hand, evidently $\CK$ is smooth over $\BA_{(r',r'')}$, and we
have an evident retraction of $\BA_{(r',r'')}$ on $S_F$ (see Fig. 1).
Therefore, $R\Gamma(\BA_{(r',r'')}-Y,S_F;\CK)=0$ which proves the claim.

\begin{picture}(20,8)(-10,-4)

\put(0,0){\circle{0.2}}
\put(0,0){\oval(6,6)}
\put(-1.5,3.3){$S_{r''}$}

\put(-4,0){\line(1,0){8}}

\put(0,3){\line(0,-1){2}}
\put(0,-3){\line(0,1){2}}
\put(-0.4,1.8){$Y$}

\put(0,1){\circle*{0.2}}
\put(0,0.5){$i\epsilon w_+$}
\put(0,-1){\circle*{0.2}}

\put(2,0){\circle*{0.2}}
\put(2,-0.5){$w_+$}
\put(-2,0){\circle*{0.2}}
\put(-2,-0.5){$w_-$}

\put(0,0){\oval(3,3)}
\put(-1.5,1.6){$S_{r'}$}

\put(0,0){\oval(5,5)[tr]}
\put(0.5,2.5){\circle*{0.2}}
\put(0.5,2){$z$}
\put(0.5,2.5){\vector(1,0){1}}

\put(-0.5,-4){Fig. 1}

\end{picture}

A clockwise rotation by $\pi/2$ induces an isomorphism
$$
R\Gamma(\BA_{\leq r'},Y\cap \BA_{\leq r'};D\CK)\cong
R\Gamma(\BA_{\leq r'},\epsilon\cdot D_F^{opp};D\CK),
$$
and the last complex is isomorphic to $\Phi_F(D\CK)$ by dilatation and
{}~(\ref{isom-one}).
This proves the theorem for $\Phi_F$. The statement for  functors
$\Phi_{E_{\pm}}$ is evident.

Let us return to the case of an arbitrary arrangement.

\subsection{}
\label{isom}
\begin{lem}{} Suppose that $\CH$ is central. Let $F$ be the unique $0$-
dimensional facet. The evident restriction maps induce canonical isomorphisms
$$
\Phi_{F}(\CK)\overset{(1)}{\cong} R\Gamma(\BA_{\BR},S_{F};\CK)
\overset{(2)}{\cong} R\Gamma(\BA,S_{F};\CK)
\overset{(3)}{\cong} R\Gamma(\BA,D_{F}^{opp};\CK)
\overset{(4)}{\cong} R\Gamma(\BA_{\BR},D_{F}^{opp};\CK)
$$
where $D_{F}^{opp}=\BA_{\BR}-\DO_{F}$.
\end{lem}

{\bf Proof.} Let us fix a coordinate system with the origin at $F$, and
hence a metric on $\BA$.
For $\epsilon>0$ let $U_{\epsilon}\subset \BA_{\BR}$ denote
the set of points $x\in \BA_{\BR}$ having distance $<\epsilon$ from
$D_{F}$.

It follows from ~\ref{limit} that
$$
\Phi_{F}(\CK)=R\Gamma(D_{F},S_{F};\CK)\cong
\dirlim_{\epsilon}\ R\Gamma(U_{\epsilon},S_{F};\CK).
$$
On the other hand, from ~\ref{conicl} it follows that restriction maps
$$
R\Gamma(\BA_{\BR},S_{F};\CK)\lra R\Gamma(U_{\epsilon},S_{F};\CK)
$$
are isomorphisms. This establishes an isomorphism (1). To prove (3), one
remarks that its cone is acyclic.

The other isomorphisms are proven by the similar arguments. We leave
the proof to the reader. $\Box$

\subsection{Proof of ~\ref{dual}}
\label{mainconst} First let us suppose that $\CH$ is central
and $F$ is its $0$-dimensional facet. We fix a coordinate system in $\BA$ as
in the proof of the lemma above.
We have a decomposition $\BA=\BA_{\BR}\oplus i\cdot \BA_{\BR}$. We will denote
by $\Re,\Im:\BA\lra\BA_{\BR}$ the evident projections.

Let
$\BA_{\leq r}$, etc. have the meaning similar to the one-dimensional case
above.
Let us denote $j:=j_{\BA-S_F}$. We proceed as in one-dimensional case.

Let us choose positive numbers $r',r'',\ \epsilon,$ such that
$$
\epsilon D_F\subset \BA_{<r'}\subset \DO_F\subset D_F\subset
\BA_{< r''}
$$
Let us introduce a subspace
$$
Y=\epsilon i\cdot D_F^{opp}
$$
We have isomorphisms
$$
D\Phi_{F}(\CK)\cong DR\Gamma(\BA,S_{F};\CK)
$$
(by Lemma ~\ref{isom})
$$
\cong R\Gamma_c(\BA; j_{*}j^*D\CK)
\cong R\Gamma(\BA,\BA_{\geq r''};j_{*}j^*D\CK)\cong
R\Gamma(\BA,Y\cup \BA_{\geq r''};j_{*}j^*D\CK)
$$
(homotopy).
Consider the restriction map
\begin{equation}
res: R\Gamma(\BA,Y\cup \BA_{\geq r''};j_*j^*D\CK)\lra
R\Gamma(\BA_{\leq r'},Y\cap \BA_{\leq r'};D\CK)
\end{equation}
$\Cone (res)$ is isomorphic to
\begin{eqnarray}
R\Gamma(\BA,\BA_{\leq r'}\cup\BA_{\geq r''}\cup Y;j_*j^*D\CK)=
R\Gamma_c(\BA_{< r''},\BA_{\leq r'}\cup Y;j_*j^*D\CK)\cong\nonumber\\
\cong DR\Gamma(\BA_{< r''}-(\BA_{\leq r'}\cup Y);j_!j^*\CK)\nonumber
\end{eqnarray}
We have by definition
$$
R\Gamma(\BA_{< r''}-(\BA_{\leq r'}\cup Y);j_!j^*\CK)=
R\Gamma(\BA_{(r',r'')}-Y,S_F;\CK)
$$

\subsubsection{}
\label{acycl} {\bf Lemma.\ }{\em Set $\BA'_{\BR}:=\BA_{\BR}-\{ 0\}.$ We have
$R\Gamma(\BA-i\cdot \BA_{\BR},\BA'_{\BR};\CK)=0$.}

{\bf Proof.}
We have to prove that the restriction
map
\begin{equation}
\label{puncture}
R\Gamma(\BA-i\cdot \BA_{\BR};\CK)\lra R\Gamma(\BA'_{\BR};\CK)
\end{equation}
is an isomorphism. Consider projection to the real part
$$
\pi: \BA-i\cdot \BA_{\BR}\lra\BA'_{\BR}.
$$
Evidently,
$$
R\Gamma(\BA-i\cdot \BA_{\BR};\CK)=R\Gamma(\BA'_{\BR},\pi_*\CK).
$$
On the other hand, since $\CK$ is conic along fibers of $\pi$, we have
$$
\pi_*\CK\iso e^*\CK
$$
where $e:\BA'_{\BR}\lra \BA-i\cdot \BA_{\BR}$ denotes the embedding
(cf. \ref{conic}). This implies our lemma. $\Box$

It follows easily that $\Cone(res)$ is acyclic, therefore the map $res$ is an
isomorphism. In other words, we have constructed an isomorphism
$$
D\Phi_F(\CK)\iso R\Gamma(\BA_{\leq r'},Y\cap \BA_{\leq r'};D\CK).
$$
%%%metka
A clockwise rotation by $\pi/2$ induces an isomorphism
$$
R\Gamma(\BA_{\leq r'},Y\cap \BA_{\leq r'};D\CK)\cong
R\Gamma(\BA_{\leq r'},\epsilon\cdot D_F^{opp};D\CK),
$$
and the last complex is isomorphic to $\Phi_F(D\CK)$ by dilatation and
{}~\ref{isom} (3).

This proves the theorem for $\Phi_F$. Note that we have constructed an
explicit isomorphism.

\subsubsection{}
\label{transv} If $F$ is an arbitrary facet, we consider a transversal
slice $L$ as in ~\ref{slice}. By the results of {\em loc.cit.}, and
the above proven case, we have natural isomorphisms
$$
D\Phi_F(\CK)\iso D\Phi_{\{^Fw\} }(\CK_L)\iso \Phi_{\{^Fw\} }(D\CK_L)\iso
\Phi_{\{^Fw\}}((D\CK)_L)\iso \Phi_F(D\CK).
$$
This proves the theorem. $\Box$

\subsection{}
\label{exact}
\begin{thm}{} For every $\CM\in\CM(\BA;\CS_{\CH})$ and every facet $F$
we have $H^i(\Phi_{F}(\CM))=0$ for all $i\neq 0$.
\end{thm}

In other words, functors $\Phi_F$ are $t$-exact with respect to the middle
perversity.

{\bf Proof.} First let us suppose that $\CH$ is central and $F$ is its
$0$-dimensional facet. Let us prove that $\Phi_F$ is right exact,
that is, $H^i(\Phi_F(\CM))=0$ for $i>0$ and every $\CM$ as above.

In fact, we know that

\subsubsection{}
\label{estim} {\em if $S$ is any stratum of $\CS_{\CH}$ of dimension $p$
then $\CM|_S\in\CD^{\leq -p}(S)$}

by the condition of perversity. In particular, $R\Gamma(\BA;\CM)=
i_F^*\CM\in\CD^{\leq 0}(\{ pt\})$.

On the other hand, one deduces from ~\ref{estim} that
$R\Gamma(S_F;\CM)\in\CD^{\leq -1}(\{ pt\})$. In fact, by definition
$S_F$ is a union of certain simplices $\Delta$, all of whose edges
lie in strata of positive dimension. This implies that
$R\Gamma(\Delta;\CM|_{\Delta})\in \CD^{\leq -1}(\{ pt\})$, and one concludes
by Mayer-Vietoris argument, using similar estimates for intersections
of simplices.

Consequently we have
$$
\Phi_{F}(\CM)\cong R\Gamma(\BA,S_F;\CM)\in\CD^{\leq 0}(\{ pt\}),
$$
as was claimed.

On the other hand by Duality theorem ~\ref{dual} we have an opposite
inequality, which proves that $\Phi_F$ is exact in our case.

The case of an arbitrary facet is reduced immediately to the central one by
noting that an operation of the restriction to a transversal slice
composed with a shift by its codimension is $t$-exact and using
{}~(\ref{sliceiso}). The theorem is proved. $\Box$

\subsection{} By the above theorem, the restriction of functors
$\Phi_F$ to the abelian subcategory $\CM(\BA,\CS_{\CH})$ lands
in subcategory $\Vect\subset\CD(\{ pt\})$.

In other words, we get exact functors
\begin{equation}
\label{phi}
\Phi_F:\CM(\BA,\CS_{\CH})\lra\Vect
\end{equation}
These functors commute with Verdier duality.

We will also use the notation $\CM_F$ for $\Phi_F(\CM)$.

\subsection{Canonical and variation maps} Suppose we have a facet $E$.
Let us denote by $\Fac^1(E)$ the set of all facets $F$ such that
$E<F$, $\dim\ F=\dim\ E+1$. We have
\begin{equation}
\label{unican}
S_E=\bigcup_{F\in\Fac^1(E)}D_F
\end{equation}

Suppose we have $\CK\in\CD(\BA,\CS_{\CH})$.

\subsubsection{}
\label{addcan} {\bf Lemma.} {\em We have a natural isomorphism
$$
R\Gamma(S_E,\bigcup_{F\in\Fac^1(E)}S_F;\CK)\cong
\oplus_{F\in\Fac^1{E}}R\Gamma(D_F,S_F;\CK)
$$}

{\bf Proof.} Note that $S_E-\bigcup_{F\in\Fac^1(E)}S_F=
\bigcup_{F\in\Fac^1(E)}\DO_F$ (disjoint union). The claim follows now from the
Poincar\'{e} duality. $\Box$

Therefore, for any $F\in\Fac^1(E)$ we get a natural inclusion map
\begin{equation}
\label{projcan}
i^F_E: R\Gamma(D_F,S_F;\CK)\hra R\Gamma(S_E,\bigcup_{F'\in\Fac^1(E)}S_{F'};\CK)
\end{equation}

Let us define a map
$$
u^F_E(\CK):\Phi_F(\CK)\lra\Phi_E(\CK)
$$
as a composition
$$
R\Gamma(D_F,S_F;\CK)[-p]\overset{i^F_E}{\lra}
R\Gamma(S_E,\bigcup_{F'\in\Fac^1(E)}S_{F'};\CK)[-p]\lra
R\Gamma(S_E;\CK)[-p]\lra
R\Gamma(D_E,S_E)[-p+1]
$$
where the last arrow is the coboundary map for the couple $(S_E,D_E)$,
and the second one is evident.

This way we get a natural transormation
\begin{equation}
\label{can}
u^F_E:\Phi_F\lra \Phi_E
\end{equation}
which will be called a {\em canonical map}.

We define a {\em variation map}
\begin{equation}
\label{var}
v^E_F:\Phi_E\lra\Phi_F
\end{equation}
as follows. By definition, $v^E_F(\CK)$ is the map dual to
the composition
$$
D\Phi_F(\CK)\iso\Phi_F(D\CK)\overset{u^F_E(D\CK)}{\lra}\Phi_E(D\CK)
\iso D\Phi_E(\CK).
$$

\subsection{Lemma}
\label{trans} {\em Suppose we have $4$ facets $A,B_1,B_2,C$ such that
$A<B_1<C,\ A<B_2<C$ and $\dim\ A=\dim\ B_i-1=\dim\ C-2$ (see Fig. 2). Then
$$
u^{B_1}_A\circ u^C_{B_1}=-u^{B_2}_A\circ u^C_{B_2}
$$
and
$$
v^{B_1}_C\circ v^A_{B_1}=-v^{B_2}_C\circ v^A_{B_2}.
$$}

For a proof, see below, ~\ref{prooftr}.

\begin{picture}(20,8)(-10,-4)

\put(0,-2){\line(1,1){4}}
\put(2.5,0){$B_2$}
\put(0,-2){\line(-1,1){4}}
\put(-3,0){$B_1$}

\put(0,-2){\circle*{0.2}}
\put(0,-2.7){$A$}

\put(0,3){$C$}

\put(-0.5,-4){Fig. 2}

\end{picture}

\subsection{Cochain complexes}
For each integer $p$, $0\leq p\leq N$, and $\CM\in\CM(\BA,\CS_{\CH})$
introduce vector spaces
\begin{equation}
C^{-p}_{\CH}(\BA;\CM)=\oplus_{F:\dim F=p}\ \CM_F
\end{equation}
For $i>0$ or $i<-N$ set $C^i_{\CH}(\BA;\CM)=0$.

Define operators
$$
d: C^{-p}_{\CH}(\BA;\CM)\lra C^{-p+1}_{\CH}(\BA;\CM)
$$
having components $u^F_E$.

\subsubsection{}
\label{nilp} {\bf Lemma.} {\em $d^2=0$}.

{\bf Proof.} Let us denote $X_p:=\bigcup_{F:\dim F=p}D_F$. We have
$X_p\supset X_{p+1}$. Evident embeddings
of couples $(D_F,S_F)\hra (X_p,X_{p+1})$ induce maps
$$
R\Gamma(X_p,X_{p+1}; \CM)\lra \oplus_{F:\dim F=p}\ R\Gamma(D_F,S_F;\CM)
$$
which are easily seen to be isomorphisms. Thus, we can identify
$C^{-p}(\BA;\CM)$ with $R\Gamma(X_p,X_{p+1};\CM)$. In these terms,
$d$ is a boundary homomorphism for the triple $X_p\supset X_{p+1}
\supset X_{p+2}$. After this description, the equality $d^2=0$ is a general
fact from homological algebra. $\Box$

\subsubsection{}
\label{prooftr} {\bf Proof of ~\ref{trans}.} The above lemma is equivalent
to the statement of ~\ref{trans} about maps $u$, which is thus proven.
The claim for variation maps follows by duality. $\Box$

This way we get a complex $C^{\bullet}_{\CH}(\BA;\CM)$ lying in
degrees from $-N$ to $0$. It will be called the {\em cochain
complex} of our arrangement $\CH_{\BR}$ with coefficients in $\CM$.

\subsection{Theorem}
\label{rgamma} {\em (i) A functor
$$
\CM\mapsto C^{\bullet}_{\CH}(\BA;\CM)
$$
is an exact functor from $\CM(\BA;\CS_{\CH})$ to the category of complexes
of vector spaces.

(ii) We have a canonical natural isomorphism in
$\CD(\{ pt\})$
$$
C^{\bullet}_{\CH}(\BA;\CM)\iso R\Gamma(\BA;\CM)
$$}

{\bf Proof.} (i) is obvious from the exactness of functors $\Phi_F$,
cf. Thm. ~\ref{exact}.
To prove (ii), let us consider the filtration
$$
\BA\supset X_0\supset X_1\supset\ldots X_N\supset 0.
$$
It follows easily from homotopy argument (cf. ~\ref{limit}, ~\ref{conic})
that the restriction
$$
R\Gamma(\BA;\CM)\lra R\Gamma(X_0;\CM)
$$
is an isomorphism. On the other hand, a "Cousin" interpretation of
$C^{\bullet}_{\CH}(\BA;\CM)$ given in the proof of Lemma ~\ref{nilp},
shows that one has a canonical isomorphism
$R\Gamma(X_0;\CM)\iso C^{\bullet}_{\CH}(\BA;\CM)$. $\Box$

\section{Computations for standard sheaves}

\subsection{}
\label{poinc} Suppose we have a connected locally simply connected
topological space $X$ and a subspace $Y\subset X$ such that each connected
component of $Y$ is simply connected.
Recall that a {\em groupoid} is a category all of whose morphisms are
isomorphisms. Let us define a {\em Poincar\'{e} groupoid}
 $\pi_1(X;Y)$ as follows.

We set $\Ob \pi_1(X;Y)=\pi_0(Y)$. To define morphisms, let us choose
a point $y_i$ on each connected component $Y_i\subset Y$. By definition,
for two connected components $Y_i$ and $Y_j$, the set of homomorphisms
$\Hom_{\pi_1(X,Y)}(Y_i,Y_j)$ is the set of all homotopy classes of paths
in $X$ starting at $y_i$ and ending at $y_j$.

A different choice of points $y_i$ gives a canonically isomorphic
groupoid. If $Y$ is reduced to one point we come back to a usual definition of
the fundamental group.

Given a local system $\CL$ on $X$, we may assign to it a "fiber" functor
$$
F_{\CL}:\pi_1(X;Y)\lra\Vect,
$$
carrying $Y_i$ to the fiber $\CL_{y_i}$. This way we get an equivalence
of the category of local systems on $X$ and the category of functors
$\pi_1(X,Y)\lra\Vect$.

\subsection{} Return to the situation of the previous section. It is known
(cf. ~\cite{br}) that the homology group $H_1(\BAO_{\CH};\BZ)$ is a free
abelian group with a basis consisting of classes of small loops around
hyperplanes $H\in\CH$. Consequently, for each map
\begin{equation}
\label{monodr}
\bq: \CH\lra\BC^*
\end{equation}
there exists a one-dimensional local system $\CL(\bq)$ whose monodromy
around $H\in \CH$ is equal to $\bq(H)$. Such a
local system is unique up to a non-unique isomorphism.

Let us construct such a local system explicitely, using a language of the
previous subsection.

\subsection{} From now on we fix
a real equation for each $H\in\CH$, i.e. a linear function
$f_H:\BA_{\BR}\lra\BR$
such that $H_{\BR}=f^{-1}(0)$. We will denote also by $f_H$ the induced
function $\BA\lra\BC$.

The hyperplane $H_{\BR}$ divides $\BA_{\BR}$ into two halfspaces:
$\BA^+_{\BR,H}=\{ x\in\BA_{\BR}|f_H(x)>0\}$ and
{}~{$\BA^-_{\BR,H}=\{ x\in\BA_{\BR}|f_H(x)<0\}$}.

Let $F\subset H_{\BR}$ be a facet of dimension $N-1$.
We have two chambers $F_{\pm}$ adjacent to $F$, where
$F_{\pm}\subset\BA_{\BR,H}^{\pm}$. Pick a point $w\in F$.
Let us choose a real affine line $l_{\BR}\subset\BA_{\BR}$ transversal
to $H_{\BR}$ and passing through $w$. Let $l$ denote its complexification.

The function $f_H$ induces isomorphism $l\iso\BC$, and
$f_H^{-1}(\BR)\cap l=l_{\BR}$. Let us pick a real $\epsilon>0$ such that two
points
$f_H^{-1}(\pm\epsilon)\cap l_{\BR}$ lie in $F_{\pm}$ respectively. Denote
these points by $w_{\pm}$.

\begin{picture}(20,8)(-10,-4)

\put(-1,-3){\line(1,1){5}}
\put(-1.5,-3.5){$H_{\BR}$}
\put(0.3,-1.2){$F$}

\put(1.5,-0.5){\circle*{0.2}}
\put(1.2,-1){$w$}

\put(1.5,-0.5){\line(1,-1){1}}
\put(2,-1){\circle*{0.2}}
\put(2.3,-1){$w_+$}

\put(1.5,-0.5){\line(-1,1){1}}
\put(1,0){\circle*{0.2}}
\put(0.2,0){$w_-$}

\put(1,-3){\line(-1,1){5}}

\put(-1,0){$F_-$}
\put(3,0){$F_+$}

\put(-4,1){\line(1,0){8}}
\put(-3,1){\circle*{0.2}}
\put(3,1){\circle*{0.2}}

\put(0,-2){\circle*{0.2}}

\put(-0.5,-4){Fig. 3}

\end{picture}

Let us denote by  $\tau^+$ (resp., $\tau^-$) a counterclockwise
(resp., clockwise) path in the upper (resp., lower) halfplane connecting
$\epsilon$ with $-\epsilon$. Let us denote
$$
\tau^{\pm}_{F}=f_H^{-1}(\tau^{\pm})
$$
This way we get two well-defined homotopy classes of paths
connecting chambers $F_+$ and $F_-$. The argument $\arg\ f_H$
increases by $\mp\frac{\pi}{2}$ along $\tau^{\pm}_F$.

Note that if $H'$ is any other hyperplane of our arrangement then
$\arg f_{H'}$ gets no increase along $\tau^{\pm}_F$.

\subsection{} Now suppose we have $\bq$ as in ~(\ref{monodr}).
Note that all connected components of $\BAO_{\CH,\BR}$ --- chambers
of our arrangement --- are contractible.
Let us define a functor
$$
F(\bq^2):\pi_1(\BAO_{\CH},\BAO_{\CH,\BR})\lra\Vect_{\BC}
$$
as follows. For each chamber $C$ we set $F(\bq^2)(C)=\BC$.
For each facet $F$ of codimension $1$ which lies in a hyperplane $H$
we set
$$
F(\bq^2)(\tau^{\pm}_F)=\bq(H)^{\pm 1}
$$
It follows from the above remark on the structure of $H_1(\BAO_{\CH})$ that
we get a correctly defined functor.

The corresponding abelian local system over $\BA_{\CH}$ will be denoted
$\CL(\bq^2)$; it has a monodromy $\bq (H)^2$ around $H\in\CH$.
If all numbers $\bq(H)$ belong to some subfield $B\subset\BC$ then
the same construction gives a local system of $B$-vector spaces.

\subsection{} From now on untill the end of this section we fix a map
$$
\bq:\CH\lra B^*,
$$
$B$ being a subfield of $\BC$, and denote by $\CL$ the local system
of $B$-vector spaces $\CL(\bq^2)$ constructed above. We denote by
$\CL^{-1}$ the dual local system $\CL(\bq^{-2})$.

Let $j:\BAO_{\CH}\lra\BA$ denote an open embedding. For $?$ equal to
one of the symbols $!,*$ or $!*$,
let us consider perverse sheaves $\CL_{?}:=j_?\CL[N]$. They belong to
$\CM(\BA;\CS_{\CH})$; these sheaves will be called {\em standard extensions}
of $\CL$, or simply {\em standard sheaves}. Note that
\begin{equation}
\label{dualstand}
D\CL_!\cong\CL^{-1}_*;\ D\CL_*\cong\CL^{-1}_!
\end{equation}
We have a canonical map
\begin{equation}
\label{map!*}
m: \CL_!\lra\CL_*,
\end{equation}
and by definition $\CL_{!*}$ coincides with its image.

Our aim in this section will be to compute explicitely the cochain
complexes of standard sheaves.

\subsection{Orientations}
\label{coor} Let $F$ be a facet which is a
connected component of a real edge $L_F$. Consider a linear space
$L^{\bot}_F=\BA_{\BR}/L_F$. Let us define
$$
\lambda_F:=H^0(L^{\bot}_F;\Or_{L^{\bot}_F}),
$$
it is a free abelian group of rank $1$. To choose an orientation
of $L^{\bot}_F$ (as a real vector space) is the same as to choose
a basis vector in $\lambda_F$.
We will call an orientation of $L^{\bot}_F$ a {\em coorientation}
of $F$.

We have an evident piecewise linear homeomorphism of
$D_F=\bigcup D_{F<C}$ onto a closed disk in $L^{\bot}_F$; thus, a
coorientation of $F$ is the same as an orientation of $D_F$ (as a $C^0$-
manifold); it defines orientations of all cells $D_{F<C}$.

\subsubsection{}
\label{sign} From now on untill the end of the section, let us fix
coorientations of all facets.
Suppose we have a pair $E<F$, $\dim\ E=\dim\ F-1$. The cell $D_E$ is a part
of the boundary of $D_F$. Let us define the sign $\sgn(F,E)=\pm 1$ as follows.
Complete an orienting basis of $D_E$ by a vector directed outside
$D_F$; if we get the given orientation of $D_F$, set $\sgn(F,E)=1$,
otherwise set $\sgn(F,E)=-1$.

\subsection{Basis in $\Phi_F(\CL_!)^*$}

Let $F$ be a facet of dimension $p$. We have by definition
$$
\Phi_F(\CL_!)=H^{-p}(D_F,S_F;\CL_!)=H^{N-p}(D_F,S_F;j_!\CL)\cong
H^{N-p}(D_F,S_F\cup (_{\CH}H_{\BR}\cap D_F);j_!\CL).
$$
By Poincar\'{e} duality,
$$
H^{N-p}(D_F,S_F\cup (_{\CH}H_{\BR}\cap D_F);j_!\CL)^*\cong
H^0(D_F-(S_F\cup _{\CH}H_{\BR});\CL^{-1})
$$
(recall that we have fixed an orientation of $D_F$).
The space $D_F-(S_F\cup _{\CH}H_{\BR})$ is a disjoint union
$$
D_F-(S_F\cup _{\CH}H_{\BR})=\bigcup_{C\in \Ch (F)}\DO_{F<C}.
$$
Consequently,
$$
H^0(D_F-(S_F\cup _{\CH}H_{\BR});\CL^{-1})\cong
\oplus_{C\in \Ch (F)}H^0(\DO_{F<C};\CL^{-1}).
$$
By definition of $\CL$, we have canonical identifications
$H^0(\DO_{F<C};\CL^{-1})=B$. We will denote by $c(\CL_!)_{F<C}\in\Phi_F(\CL)^*$
the
image of $1\in H^0(\DO_{F<C};\CL^{-1})$ with respect to the embedding
$$
H^0(\DO_{F<C};\CL^{-1})\hra \Phi_F(\CL_!)^*
$$
following from the above.

Thus, classes $c(\CL_!)_{F<C},\ C\in \Ch (F)$, form a basis of
$\Phi_F(\CL_!)^*$.

\subsection{}
\label{can!} Let us describe canonical maps for
$\CL_!$. If $F<E,\ \dim\ E=\dim\ F+1$, let
$u^*:\Phi_F(\CL_!)^*\lra\Phi_E(\CL_!)^*$ denote the map dual to $u^E_F(\CL_!)$.
Let $C$ be a chamber adjacent to $F$.
Then
\begin{equation}
u^*(c(\CL_!)_{F<C})=\left\{ \begin{array}{ll}
                        \sgn(F,E)c(\CL_!)_{E<C}&\mbox{if $E<C$}\\
                          0&\mbox{otherwise}
                     \end{array}
             \right.
\end{equation}

\subsection{Basis in $\Phi_F(\CL_*)^*$} We have isomorphisms
\begin{equation}
\label{dualchains}
\Phi_F(\CL_*)\cong\Phi_F(D\CL_!^{-1})\cong\Phi_F(\CL_!^{-1})^*
\end{equation}
Hence, the defined above basis $\{ c(\CL^{-1}_!)_{F<C}\}_{C\in\Ch (F)}$
of $\Phi_F(\CL_!^{-1})^*$, gives a basis in  $\Phi_F(\CL_*)$.
We will denote by $\{ c(\CL_*)_{F<C}\}_{C\in\Ch (F)}$ the dual basis
of $\Phi_F(\CL_*)$.

\subsection{Example} Let us describe our chains explicitely in the simplest
one-dimensional case, in the setup ~\ref{one-dim}. We choose a natural
orientation on $\BA_{\BR}$.
A local system $\CL=\CL(q^2)$ is uniquely determined by one nonzero
complex number $q$. By definition, the upper (resp., lower) halfplane
halfmonodromy from  $w_+$ to $w_-$ is equal to $q$ (resp., $q^{-1}$).

\subsubsection{Basis in $\Phi_F(\CL_!)^*$} The space $\Phi_F(\CL_!)^*$
admits a basis consisting of two chains
$c_{\pm}=c(\CL_!)_{F<E_{\pm}}$ shown below, see Fig. 3(a).
By definition, a homology class is represented by a cell together with
a section of a local system $\CL^{-1}$ over it. The section of $\CL^{-1}$
over $c_+$ (resp., $c_-$) takes value $1$ over $w_+$ (resp., $w_-$).

%%%%%%%%%%%%%%% picture

\begin{picture}(20,8)(-10,-4)

\put(-5,0){\circle{0.2}}
\put(-5,0){\oval(6,6)}
\put(-6.5,3.3){$S_{r''}$}

\put(-3,0){\circle*{0.2}}
\put(-3,-0.5){$w_+$}
\put(-7,0){\circle*{0.2}}
\put(-7,-0.5){$w_-$}

\put(-7,0){\line(1,0){4}}
\put(-7,0){\vector(1,0){1.5}}
\put(-5,0){\vector(1,0){1.5}}

\put(-6.7,0.3){$c_-$}
\put(-3.5,0.3){$c_+$}

%%%% dual chains

\put(-6,-3){\line(0,1){6}}
\put(-6,1){\vector(0,1){1}}
\put(-6,-3){\circle*{0.2}}
\put(-6,3){\circle*{0.2}}
\put(-5.8,2){$Dc_-$}

\put(-4,-3){\line(0,1){6}}
\put(-4,1){\vector(0,1){1}}
\put(-4,-3){\circle*{0.2}}
\put(-4,3){\circle*{0.2}}
\put(-3.8,2){$Dc_+$}

\put(-5.5,-3.5){(a)}

%%%%%%%%%%%%%%%  deformed chains

\put(5,0){\circle{0.2}}
\put(5,0){\oval(6,6)}
\put(3.5,3.3){$S_{r''}$}

\put(5,3){\line(0,-1){2}}
\put(5,-3){\line(0,1){2}}
\put(4.6,1.8){$Y$}

\put(5,1){\circle*{0.2}}
\put(5,-1){\circle*{0.2}}

\put(7,0){\circle*{0.2}}
\put(7,-0.5){$w_+$}
\put(3,0){\circle*{0.2}}
\put(3,-0.5){$w_-$}

\put(3,0){\line(1,0){4}}
\put(3,0){\vector(1,0){1.5}}
\put(5,0){\vector(1,0){1.5}}

\put(-6.7,0.3){$c_-$}
\put(-3.5,0.3){$c_+$}

\put(5,0){\oval(2,3)}
\put(6,0){\vector(0,1){0.5}}
\put(6.2,0.5){$\tilde{D}c_+$}
\put(4,0){\vector(0,1){0.5}}
\put(3,0.5){$\tilde{D}c_-$}
\put(5,1.5){\circle*{0.2}}
\put(5,-1.5){\circle*{0.2}}

\put(4.5,-3.5){(b)}

\put(-0.5,-4){Fig. 3}

\end{picture}

\subsubsection{Basis in $\Phi_F(\CL_*)^*$}
Let us adopt notations of ~\ref{mainconst}, with $\CK=\CL^{-1}_!$.
It is easy to find the basis $\{ Dc_+,Dc_-\}$ of the dual space
$$
\Phi_F(\CL_!^{-1})\cong H^0(\BA,\BA_{\geq r''};j^*j_*D\CL_!^{-1})^*=
H^1(\BA-\{ w_+,w_-\},\BA_{\geq r''};\CL)^*
$$
dual to $\{ c(\CL^{-1})_{F<E_+},c(\CL^{-1})_{F<E_-}\}$.
Namely, $Dc_{\pm}$ is represented by the relative $1$-chain
$$
\{ \pm\frac{1}{2}+y\cdot i|\ -\sqrt{(r'')^2-\frac{1}{4}}\leq y
\leq \sqrt{(r'')^2-\frac{1}{4}}\},
$$
with evident sections of $\CL^{-1}$ over them, see Fig. 3(a).

Next, one has to deform these chains to chains $\tilde{D}c_{\pm}$ with
their ends on $Y$, as in Fig. 3(b). Finally, one has to make a clockwise
rotation of the picture by $\pi/2$. As a result, we arrive at the
following two chains $c^*_{\pm}$ forming a basis of $\Phi_F(\CL_*)^*$:

%%%%%%%% picture

\begin{picture}(20,8)(-10,-4)

\put(0,0){\circle{0.2}}

\put(2,0){\circle*{0.2}}
\put(2.2,-0.5){$w_+$}
\put(-2,0){\circle*{0.2}}
\put(-1.8,-0.5){$w_-$}

\put(0,0){\oval(4,4)}
\put(0,2){\vector(1,0){0.5}}
\put(0,2.3){$c^*_-$}
\put(0,-2){\vector(1,0){0.5}}
\put(0,-2.5){$c^*_+$}

\put(-2,0){\line(1,0){4}}
\put(-2,0){\vector(1,0){1.5}}
\put(0,0){\vector(1,0){1.5}}

\put(-1,0.3){$c_-$}
\put(0.8,0.3){$c_+$}

\put(-0.5,-4){Fig. 4}

\end{picture}

The section of $\CL^{-1}$ over $c^*_+$ (resp., $c_-$) has value $1$ at
$w_+$ (resp., $w_-$).

It follows from this description that the natural map
\begin{equation}
\label{m!*}
m:\Phi_F(\CL_*)^*\lra\Phi_F(\CL_!)^*
\end{equation}
is given by the formulas
\begin{equation}
\label{f!*}
m(c^*_+)=c_++qc_-;\ m(c^*_-)=qc_++c_-
\end{equation}

By definition, spaces $\Phi_{E_{\pm}}(\CL_?)^*$ may be identified
with fibers $\CL_{w_{\pm}}$ respectively, for both $?=!$ and $?=*$,
and hence with $B$. Let us denote by $c_{w_{\pm}}$ and
$c^*_{w_{\pm}}$ the generators corresponding to $1\in B$.

It follows from the above description that the canonical maps $u^*$ are given
by the formulas
\begin{equation}
\label{fu!}
u^*(c_+)=c_{w_+};\ u^*(c_-)=-c_{w_-};
\end{equation}
and
\begin{equation}
\label{fu*}
u^*(c_+^*)=c_{w_+}^*-qc_{w_-}^*;\
u^*(c_-^*)=qc_{w_+}^*-c_{w_-}^*;\
\end{equation}
Let us compute variation maps. To get them for the sheaf $\CL_!$, we should
by definition replace $q$ by $q^{-1}$ in ~(\ref{fu*}) and take the conjugate
map:
\begin{equation}
\label{fv!}
v^*(c_{w_+})=c_+ + q^{-1}c_-;\
v^*(c_{w_-})=-q^{-1}c_+-c_-
\end{equation}
To compute $v^*$ for $\CL_*$, note that the basis in
$$
H^0(\BA,\{ w_+,w_-\};\CL_!^{-1})^*=H_1(\BA,\{w_+,w_-,0\};\CL)
$$
dual to $\{ c^*_+,c^*_-\}$, is $\{ q^{-1}\tc_-,q^{-1}\tc_+\}$ {\em (sic!)}
where $\tc_{\pm}$ denote the chains defined in the same way as $c_{\pm}$,
with $\CL$ replaced by $\CL^{-1}$. From this remark it follows that
\begin{equation}
\label{fv*}
v^*(c_{w_+}^*)=q^{-1}c_-^*;\
v^*(c_{w_-}^*)=-q^{-1}c_+^*
\end{equation}

\subsection{} Let us return to the case of an arbitrary arrangement.
Let us say that a hyperplane $H\in\CH$ {\em separates} two chambers $C, C'$
if they lie in different halfspaces with respect to $H_{\BR}$.
Let us define numbers
\begin{equation}
\label{qsep}
\bq(C,C')=\prod\bq(H),
\end{equation}
the product over all hyperplanes $H\in \CH$ separating $C$ and $C'$.
In particular, $\bq(C,C)=1$.

\subsection{Lemma.} {\em Let $F$ be a face, $C\in\Ch(F)$. The canonical mapping
$$
m:\Phi_F(\CL_*)^*\lra\Phi_F(\CL_!)^*
$$
is given by the formula
\begin{equation}
\label{sform}
m(c(\CL_*)_{F<C})=\sum_{C'\in\Ch(F)}\bq(C,C')c(\CL_!)_{F<C'}
\end{equation}}

\subsubsection{} Since $\Phi_F(\CL_*)$ is dual to $\Phi_F(\CL_!)$, we may
view $m$ as  a {\em bilinear form} on $\Phi_F(\CL_!)$. By ~(\ref{sform})
it is {\em symmetric}.

{\bf Proof} of lemma. We generalize the argument of the previous example.
First consider the case of zero-dimensional $F$.
Given a chain $c(\CL_!^{-1})_{F<C}$, the corresponding dual chain may be
taken as
$$
Dc_{F<C}= \epsilon\cdot\ ^Cw\oplus i\cdot\BA_{\BR},
$$
were $\epsilon$ is a sufficiently small positive real.
Next, to get the dual chain $c(\CL_*)_{F<C}$, we should make a deformation
similar to the above one, and a rotation by $\frac{\pi}{2}$. It is convenient
to make the rotation first. After the rotation, we get a chain
$\BA_{\BR} -\epsilon i\cdot\ ^Cw$. The value of
$m(c(\CL_*)_{F<C})$ is given by the projection of this chain to $\BA_{\BR}$.

The coefficient at $c(\CL_!)_{F<C'}$ is given by the monodromy of $\CL^{-1}$
along the following path from $C$ to $C'$. First, go "down" from $^Cw$
to $-\epsilon i\cdot\ ^Cw$; next, travel in
$\BA_{\BR} -\epsilon i\cdot\ ^Cw$ along the straight line from
$-\epsilon i\cdot\ ^Cw$ to $-\epsilon i\cdot ^{C'}w$, and then go "up" to
$^{C'}w$. Each time we are passing under a hyperplane $H_{\BR}$
separating $C$ and $C'$, we gain a factor $\bq(H)$. This gives
desired coefficient for the case $\dim\ F=0$.

For an arbitrary $F$ we use the same argument by considering the intersection
of our picture with a transversal slice.   $\Box$

\subsection{} Let $E$ be a facet which is a component of a
real edge $L_{E,\BR}$; as usually $L$ will denote the complexification.
Let $\CH_L\subset \CH$ be a subset consisting of all hyperplanes
containg $L$. If we assign to a chamber $C\in\Ch(E)$ a unique chamber of the
subarrangement $\CH_L$ comtaining $C$, we get a bijection of $\Ch(E)$ with
the set of {\em all} chambers of $\CH_L$.

Let $F<E$ be another facet. Each chamber $C\in\Ch(F)$ is contained in a unique
chamber of $\CH_L$. Taking into account a previous bijection, we get a mapping
\begin{equation}
\label{pi}
\pi^F_E:\Ch(F)\lra\Ch(E)
\end{equation}

\subsection{Lemma.}{\em Let $F$ be a facet, $C\in\Ch(F)$. We have
\begin{equation}
\label{u*}
u^*(c(\CL_*)_{F<C})=\sum\sgn(F,E)\bq(C,\pi^F_E(C))c(\CL_*)_{E<\pi^F_E(C)},
\end{equation}
the summation over all facets $E$ such that $F<E$ and $\dim\ E=\dim\ F+1$.}

(Signs $\sgn(F,E)$ have been defined in ~\ref{sign}.)

{\bf Proof.} Again, the crucial case is $\dim\ F=0$ --- the case of arbitrary
dimension is treated using a transversal slice. So, let us suppose that
$F$ is zero-dimensional. In order to compute the coefficient of
$u^*(c(\CL_*)_{F<C})$ at $c(\CL^*)_{E<C'}$ where $E$ is a one-dimensional
facet adjacent to $F$ and $C'\in\Ch(E)$, we have to do the following.

Consider the intersection of a real affine subspace
$\BA_{\BR}-\epsilon\cdot i\cdot\ ^Cw$ (as in the proof of the previous lemma)
with a complex hyperplane $M_E$ passing through $^Ew$ and transversal
to $E$. The intersection will be homotopic to a certain chain
$c(\CL_*)_{E<C''}$ where $C''$ is easily seen to be equal to $\pi^F_E(C)$,
and the coefficient is obtained by the same rule as described
in the previous proof. The sign will appear in accordance with
compatibility of orientations of $D_F$ and $D_E$. $\Box$

\subsection{} Let us collect our results. Let us denote by
$\{ b(\CL_?)_{F<C}\}_{C\in\Ch(F)}$ the basis in $\Phi_F(\CL_?)$ dual
to $\{ c(\CL_?)_{F<C}\}$, where $?=!$ or $*$.

\subsection{Theorem.}{\em (i) The complex $C^{\bullet}_{\CH}(\BA;\CL_!)$
is described as follows. For each $p,\ 0\leq p\leq N$, the space
$C^{-p}_{\CH}(\BA;\CL_!)$ admits a basis consisting of all cochains
$b(\CL_!)_{F<C}$ where $F$ runs through all facets of $\CH_{\BR}$ of dimension
$p$, and $C$ through $\Ch(F)$. The differential
$$
d:C^{-p}_{\CH}(\BA;\CL_!)\lra C^{-p+1}_{\CH}(\BA;\CL_!)
$$
is given by the formula
\begin{equation}
\label{dl!}
d(b(L_!)_{F<C})=\sum_{E:E<F,\ \dim\ E=\dim\ F-1} \sgn(E,F)b(L_!)_{E<C}
\end{equation}

(ii) The complex $C^{\bullet}_{\CH}(\BA;\CL_*)$
is described as follows. For each $p,\ 0\leq p\leq N$, the space
$C^{-p}_{\CH}(\BA;\CL_*)$ admits a basis consisting of all cochains
$b(\CL_*)_{F<C}$ where $F$ runs through all facets of $\CH_{\BR}$ of dimension
$p$, and $C$ through $\Ch(F)$. The differential
$$
d:C^{-p}_{\CH}(\BA;\CL_*)\lra C^{-p+1}_{\CH}(\BA;\CL_*)
$$
is given by the formula
\begin{equation}
\label{dl*}
d(b(L_*)_{F<C})=\sum \sgn(E,F)\bq(C,C')b(L_*)_{E<C'},
\end{equation}
the summation over all facets $E<F$ such that $\dim\ E=\dim\ F-1$ and
all chambers $C'\in\Ch(E)$ such that $\pi^E_F(C')=C$.

(iii) The natural map of complexes
\begin{equation}
\label{mapc}
m:C^{\bullet}_{\CH}(\BA;\CL_!)\lra C^{\bullet}_{\CH}(\BA;\CL_*)
\end{equation}
induced by the canonical map $\CL_!\lra\CL_*$, is given by the formula
\begin{equation}
\label{formb}
m(b(\CL_!)_{F<C})=\sum_{C'\in\Ch(F)}\bq(C,C')b(\CL_*)_{F<C'}
\end{equation}}

All statements have already been proven.

\subsubsection{} {\bf Corollary.} {\em The complexes
$C^{\bullet}_{\CH}(\BA;\CL_!)$
and $C^{\bullet}_{\CH}(\BA;\CL_*)$ described explicitely in the above theorem,
compute the relative cohomology $H^{\bullet}(\BA,\ _{\CH}H;\CL)$ and the
cohomology of the open stratum $H^{\bullet}(\BAO,\CL)$ respectively,
and the map $m$ induces the canonical map in cohomology.}

{\bf Proof.} This follows immediately from ~\ref{rgamma}. $\Box$

This corollary was proven in ~\cite{v}, Sec. 2, by a different argument.

\subsection{Theorem.}
\label{inters} {\em The complex $C^{\bullet}_{\CH}(\BA;\CL_{!*})$
is canonically isomorphic to the image of ~(\ref{mapc}).}

{\bf Proof.} This follows from the previous theorem and the exactness
of the functor $\CM\mapsto C^{\bullet}_{\CH}(\BA;\CM)$,
cf. Thm. ~\ref{rgamma} (i). $\Box$

The above description of cohomology is analogous to ~\cite{sv}, p. I,
whose results may be considered as a "quasiclassical" version
of the above computations.

%\input{biblio}

%%%%%%%%%%%%%%%%%     biblio.tex

\end{document}